\documentclass[prb,twocolumn,showpacs,floatfix,amsfonts]{revtex4}
\usepackage{graphicx,graphics,color,epsfig}
\usepackage{bm}
\usepackage{amsmath}
\usepackage{amssymb}
\usepackage{epstopdf}

\begin{document}

\title{Odd-frequency pairing in a binary mixture of bosonic and fermionic cold atoms}
\author{Ryan M. Kalas}
\author{Alexander V. Balatsky}
\author{Dmitry Mozyrsky}
\email{mozyrsky@lanl.gov}
\affiliation{Theoretical Division, Los Alamos National Laboratory, Los Alamos, NM 87545, USA}

\date{October 2, 2008}

\begin{abstract}
We study fermionic superfluidity in a boson-single-species-fermion cold
atom mixture. We argue that apart from the standard p-wave fermion pairing mediated
by the phonon field of the boson gas, the system also exhibits s-wave pairing with
the anomalous correlator being an odd function of time or frequency. We show that
such a superfluid phase can have a much higher transition temperature than
the p-wave and may exist for sufficiently strong couplings between fermions and bosons.
These conditions for odd-frequency pairing are favorable close to the value of the coupling
at which the mixture phase-separates.  We evaluate the critical temperatures
for this system and discuss the experimental realization of this 
superfluid in ultracold atomic gases.

\end{abstract}
\maketitle

\section{Introduction}
Symmetry of superfluid phases in correlated fermion liquids has been a subject
of extensive research for many decades. Two of the most famous examples of systems
exhibiting non-trivial symmetries of the order parameter are superfluid $^3$He
and high-$T_c$ cuprates, where pairing is believed to occur in p-wave triplet
and d-wave singlet channels, respectively. In 1974, in an attempt to explain the existence
of several superfluid phases in $^3$He, Berezinskii \cite{berez} had suggested that there is
yet another possibility for the triplet pairing: while the s-wave component of the
anomalous correlator $\langle\psi_\alpha({\bf r},t)\psi_\beta({\bf r}^\prime,t^\prime)\rangle$
should identically vanish at equal times due to the Pauli principle~\cite{comment1}, it may still be nonzero for $t\neq t^\prime$ (i.e., the anomalous
correlator is an odd function of time or frequency) thus giving rise to the superfluid $^3$He A-phase.
While experiments have shown that Berezinskii's conjecture is not realized in the actual
$^3$He liquid, the idea of such nonlocal-in-time pairing has attracted considerable attention
later, with the discovery of superconductivity in cuprates \cite{bal,htc} and heavy fermion compounds \cite{hf}. In particular
it has been argued that the nonlocal character of such an order parameter provides a natural resolution to
the ``paradox'' of the coexistence between the strong, but instantaneous short-range Coulomb repulsion
and Cooper pairing in these materials. More recently the odd pairing mechanism was proposed to explain the anomalous
proximity effect in superconductor-ferromagnet junctions \cite{efetov}, where superconducting
penetration length is believed to be significantly enhanced due to the formation of the triplet
odd-parity component from the standard s-wave singlet condensate.

Despite considerable interest in the subject, the question of whether an odd-frequency phase exists
in equilibrium physical systems with no external source of pairs, i.e., as a consequence of spontaneous
symmetry breaking, remains unresolved.  Previous work has suggested that such phases may be thermodynamically
unstable \cite{cox}, but the situation remains unclear.  Here we discuss the possibility that a boson-fermion mixture,
presently realizable in atomic traps, provides an example of a system where an odd-frequency
fermionic superfluid phase may exist under the appropriate conditions.

We show that due to the interaction with the phonons in the bosonic subsystem, the
fermions at sufficiently low temperatures exhibit pairing either in the p-wave channel or in the s-wave
odd-frequency channel.  A key result is that the s-wave odd-frequency pairing exists only when
the coupling $\gamma$ (to be defined below) between the phonons and the fermions exceeds a certain
threshold value $\gamma_c$. Moreover the value of $\gamma_c$ is close to the coupling strength at which
the mixture phase-separates. That is, upon an increase in the boson-fermion coupling
the phonon mode softens, thus leading to stronger attractive interaction
between the fermions, as a result of which the odd-frequency fermionic condensate can form
in the vicinity of the phase separation transition. We estimate the transition temperatures
for the system. We also point out certain
cross-correlations between the boson and fermion densities that could possibly be detected
in cold atom mixtures as a signature of the odd-frequency superfluid phase.
\section{Model}
We describe the dilute mixture of fermions and bosons with the following Hamiltonian density:
\begin{equation}
\label{ham}
H= H_B^0+ \frac{\lambda}{2}|\psi_B^\dagger \psi_B|^2 +
   H_F^0+ \lambda '\psi_B^\dagger\psi_B\psi_F^\dagger\psi_F,
\end{equation}
where $H_{B,F}^0$ denote the Hamiltonians for noninteracting bosons
and fermions, $\lambda$ and $\lambda^\prime$ are the boson-boson and boson-fermion
coupling constants. We assume that the trap confining the particles is magnetic
and therefore the fermions are fully spin polarized. As a result, due to the Pauli
principle, direct interaction between fermion atoms is negligible. For the purposes
of the present calculation we also neglect the spatial variation of the trapping potential and
assume that the local fermion and boson densities are controlled by
the chemical potentials so that
$H_{B,F}^0 = \psi_{B,F}^\dagger(-\nabla^2/2m_{B,F} - \mu_{B,F})\psi_{B,F}$.
Also, here and in the following we set $\hbar = k_B =1$.

In order to derive an effective coupling between fermions it is convenient to rewrite the bosonic
fields in terms of the density-phase variables as $\psi_B = \sqrt{\rho}\exp{(i\phi)}$. Furthermore,
writing $\rho = \rho_0+\delta\rho$, where $\rho_0$ is a constant and $\delta\rho$ contains only
nonzero frequency $\omega_n$ and wavevector ${\bf q}$ components of $\rho(\omega_n, {\bf q})$,
expanding Eq.~(\ref{ham}) up to $O(\delta\rho^2,\phi^2)$ around $\rho_0$,  and integrating out the
phase variable, we obtain the effective ``electron''-phonon model with the Matsubara action
\begin{eqnarray}
\label{seff}
&&S_{\rm eff}=\int_{-\beta/2}^{\beta/2}d\tau\int d^3{\bf r}\left({\cal L}_F^0+\lambda'\delta\rho\,\psi_F^\dagger\psi_F\right)  \\
&&+\sum_{\omega_n, {\bf q}}{1\over 2}\left({m_B\omega_n^2\over \rho_0 q^2}+\lambda+{q^2\over 4 m_B \rho_0}\right) |\delta\rho(\omega_n,{\bf q})|^2. \nonumber
\end{eqnarray}
In Eq.~(\ref{seff}) ${\cal L}_F^0$ is free-fermion Lagrangian, $\beta=1/T$ where $T$ is the temperature,
and the last term describes phonons or Bogolubov quasiparticles with dispersion
relation $\omega = c_s q(1+\xi_0^2 q^2)^{1/2}$, where the phonon speed of sound $c_s=(\lambda\rho_0/m_B)^{1/2}$ and the boson coherence
length $\xi_0=(1/4m_B\lambda\rho_0)^{1/2}$. The phonons give rise to the nonlocal fermion
density-density interaction, with vertex $-{\lambda^\prime}^2 D^0(\omega_n, {\bf q})/2$, where ${D^0}^{-1}(\omega_n, {\bf q})$ is the expression in the brackets in
the last term in Eq.~(\ref{seff}).

It should be noted that while the effective description in
Eq.~(\ref{seff}) is valid for finite values of the coupling
constant $\lambda^\prime$, it breaks down for sufficiently large
$\lambda^\prime$ or fermion density. Indeed, by evaluating the
renormalization of the phonon Green's function $D^0$ within the
second order (in $\lambda^\prime$) perturbation theory in the
particle-hole channel (i.e., accounting for a single fermion
polarization bubble in phonon self-energy), we see that
\begin{eqnarray}
\label{phonon}
{D^0}^{-1}\rightarrow D^{-1}={m_B\omega^2\over \rho_0 q^2}+\lambda (1-\gamma)+\lambda\xi^2q^2 +O(q^4),
\end{eqnarray}
where $\gamma = {\lambda^\prime}^2 q_F^2/(2\pi^2\lambda v_F)$ and $\xi^2 =\xi_0^2+\gamma/12q_F^2$; $q_F$ and $v_F=q_F/m_F$ are the Fermi momentum
and velocity respectively \cite{agd, comment3}. Thus for $\gamma\rightarrow 1$ the phonon mode
softens at small $q$, signaling that the mixture becomes
unstable against the phase separation transition.  The value of $\gamma=1$ corresponds to the line
of spinodal decomposition in the mean field analysis where the instability shows
up as a saddle point in the free energy \cite{pethick}.
While the second order perturbation theory will not be quantitatively
accurate for $\gamma$ close to $1$, the higher order renormalizations of the phonon
propagator do not alter this conclusion on the qualitative level;
their effect merely leads to a redefinition of $\gamma$ and $\xi$ \cite{eddy}.
As a result effective interaction between fermions increases as one approaches the spinodal point, diverging
for $\omega={\bf q}=0$ at $\gamma = 1$. It should be pointed out that the phase separation
is, presumably, a first order phase transition \cite{pethick,sol}, and therefore the separated phase
becomes thermodynamically more favorable before the spinodal line. However, an estimate of nucleation
rates shows that due to extremely low temperatures as well as
relatively weak interparticle interactions such nucleation processes
are exponentially slow. That is, upon an adiabatic increase of fermion-boson interaction the system will remain in the metastable mixed state up until
the absolute instability (spinodal) line, unless the boson gas
parameter is comparable to 1, see Ref.~\cite{sol}.
\section{Gap Equation}
The onset of the pairing instability corresponds to the appearance of a nonzero anomalous correlator
$F({\bf r}-{\bf r}^\prime, \tau-\tau^\prime)=i\langle {\cal T}\psi_F({\bf r},\tau)\psi_F({\bf r}^\prime,\tau^\prime)\rangle$.
The self-consistency equation can be readily obtained within the Eliashberg formalism \cite{agd}.
Following the standard procedure
we derive the linearized eigenvalue-type equation for $F(\omega, {\bf q})$:
\begin{eqnarray}
\label{gapeqn}
G^{-1}(\omega_n,{\bf q})G^{-1}(-\omega_n,{\bf q})F(\omega_n,{\bf q})= T\sum_{\omega_n', {\bf q}'}F(\omega_n',{\bf q}')\\ \nonumber
\times {{\lambda^\prime}^2\over 2}\left[D(\omega_n-\omega_n',{\bf q}-{\bf q}')-D(\omega_n+\omega_n',{\bf q}+{\bf q}')\right],
\end{eqnarray}
where $G$ is the fermion Green's function and $\omega_n=\pi T(2n+1)$ is the Matsubara frequency.
Note that the appearance of the difference in the brackets on the right hand
side (r.h.s.) of Eq.~(\ref{gapeqn}) is due to the presence of fermions with the same spin only;
had we considered the usual singlet pairing, the difference would have been removed by the
spin part of the Cooper pair wavefunction. As a result solutions to Eq.~(\ref{gapeqn}) should
satisfy the anti-symmetry property, $F(\omega_n,{\bf q})=-F(-\omega_n,-{\bf q})$, which, as expected,
rules out a possibility of the standard s-wave even-in-time pairing.

In order to solve Eq.~(\ref{gapeqn}) we must specify the Green's functions $G$ and $D$, which are
renormalized in the particle-hole channel according to the Dyson equations $G^{-1}={G^0}^{-1}-\Sigma$ and
$D^{-1}={D^0}^{-1}-\Pi$, where $\Sigma$ and $\Pi$ are the fermion and phonon self-energies, e.g., Eq.~(\ref{phonon}).
It is well known from the Fermi liquid theory \cite{agd} that near the Fermi surface the renormalized fermion Green's function $G$ can
be written as $Z[i\omega_n - v_F^\ast (q-q_F)]^{-1}$,
where $Z$ is the quasiparticle ``weight'' coefficient ($Z\leq 1$), and $v_F^\ast = Zv_F$. Then
$G^{-1}(\omega_n,{\bf q})G^{-1}(-\omega_n,{\bf q}) = [\omega_n^2+{v_F^\ast}^2(q-q_F)^2]/Z^2$.
Next let us expand $F(\omega, {\bf q})$ and $D(\omega, {\bf q})$ in orbital harmonics using
Legendre polynomials $P_l$. For reasons to be specified below we assume that  $c_s/v_F$ is small.
Then we notice that $D$ is a relatively slowly varying function of $q$ compared to $F$
(which is strongly peaked at $q=q_F$) and therefore we can set both $|{\bf q}|$ and $|{\bf q}^\prime|$
in the $D$'s on the r.h.s.~of Eq.~(\ref{gapeqn}) equal to $q_F$. As a result the $D$'s are functions of
the angle between ${\bf q}$ and ${\bf q}^\prime$ only, and with the use of the addition theorem we obtain
\begin{eqnarray}
\label{schro}
[\omega_n^2+{v_F^\ast}^2\delta q^2]{\cal F}_l(\omega_n, \delta q) = {(\lambda^\prime Z)^2\over 2} T\sum_{\omega_n^\prime}\int{q_F^2d\delta q'\over (2\pi)^2}\\ \nonumber
\times {\cal F}_l(\omega_n^\prime, \delta q')\left[{\cal D}_l(\omega_n-\omega_n')-(-1)^l {\cal D}_l(\omega_n+\omega_n')\right],
\end{eqnarray}
where ${\cal F}_l$ is the partial component of the anomalous correlator $F$ (i.e.,
with angular momentum $l$), $\delta q \equiv q-q_F$, and
${\cal D}_l = \int_{-1}^{1}d\cos{\theta} D(\omega,q_F\sqrt{2-2\cos{\theta}})P_l(\cos{\theta})$.

It is obvious that there are two types of solutions to Eq.~(\ref{schro}):
${\cal F}_l(\omega)$ with odd $l$ and even in $\omega$ and vice versa, with even $l$ and odd in $\omega$.
For both of these solutions the bracket in the r.h.s. of Eq.~(\ref{schro}) can be replaced by
$2{\cal D}_l(\omega-\omega^\prime)$ and then we note that Eq.~(\ref{schro}) can be cast in the form
${\hat H}_l{\cal F}_l=0$, where ${\hat H}_l$ is a ``Hamiltonian'' of a particle moving in a two-dimensional
external potential $V_l(\tau, x)\sim -2 {\cal D}_l(\tau)\delta(x)$ (here $\tau$ is Matsubara time and
$x$ is Fourier conjugate to $\delta q$), and ${\cal F}_l(\beta/2,
x)={\cal F}_l(-\beta/2, x)=0$.

For $\beta\rightarrow\infty$, ${\hat H}_l$ has at least one negative
eigenvalue $\epsilon_{\infty,l}$ corresponding to the bound state of
the particle. Upon an increase of $T$ (i.e., for finite $\beta$)
the ``energy'' $\epsilon_{\infty,l}$  increases, reaching $0$
for some particular value of $\beta\equiv\beta_c$. In order to
estimate the value of $\beta_c$ we note that for compact
${\cal D}_l(\tau)$ the bound state ``wavefunction'' behaves
asymptotically as $\sim\exp(-\sqrt{-\epsilon_{\infty,l}}\tau)$, and
therefore the boundary conditions, i.e., the finiteness of $\beta$,
start to intervene when $\beta\sim 1 / \sqrt{-\epsilon_{\infty,l}}$.
Moreover, since $(-\epsilon_{\infty,l})^{-1/2}$ is the only relevant
``time'' scale (in the limit when $\beta^{-1}$ is much smaller
then the cut-off frequency $\omega_c$ of ${\cal D}_l$; $\omega_c\sim c_s q_F$), the critical temperature
$T_{c,l}=1/\beta_{c,l}$ for channel $l$ is of the order of
$\sqrt{-\epsilon_{\infty,l}}$.

Evidently the ground states of ${\hat H}_l$'s describe only the
channels with odd $l$ and even $\omega$-dependence of ${\cal F}_l$'s. The
second type of solutions, even in $l$ and odd in $\tau$,
however, correspond to the first excited states of ${\hat H}_l$,
which, obviously, are odd functions of $\omega$ or $\tau$. While
the same considerations hold for the critical temperatures for these solutions as well, there is an
important distinction: in the $\beta\rightarrow\infty$ limit
the former solutions (e.g., bound ground states) always exist for the attractive $V_l$'s, the latter may
only exist when the potential $V_l$ is strong enough. Therefore even
at zero temperature the odd-frequency (Berezinskii) states exist
only if the coupling constant $\lambda^\prime$ or $\gamma$ exceeds
a certain threshold $\gamma_c$. Moreover, since at the threshold point the
state's ``wavefunction'' is delocalized and it
localizes for stronger $V_l$, the odd-frequency
phase emerges at zero temperature when $\gamma=\gamma_c$ and extends
to non-zero temperatures at stronger couplings.

It is easy to estimate this threshold coupling $\gamma_c$ and the critical temperatures
from Eqn.~(\ref{schro}); see appendix. For
the $l=0$ odd frequency case, the zero temperature
critical coupling $\gamma_c$ can be estimated
as the solution to $\gamma_c= (\alpha^2/Z^2)/\ln[1+\alpha^2/(1-\gamma_c)]$,
where $\alpha=2q_F\xi$; see appendix. The solutions to this equation are lying in the interval $1/(1+Z^2)\leq \gamma_c\leq 1$,
depending on the value of $\alpha$. In particular, for $\alpha\gg1$ one finds
$\gamma_c\approx 1 - \alpha^2\exp(-\alpha^2/Z^2)$, and so the width
of the superfluid phase (in terms of the dimensionless coupling
$\gamma$) is exponentially small. For $\alpha\sim 1$,
$1-\gamma_c$ is {\it finite} and thus the odd-superfluid phase is most favorable
when the effective boson coherence length $\xi$ is comparable with the
interfermion distance. The transition temperature can be
calculated numerically, see appendix, and the results are shown in
Fig.~1(a-c) for different values of $\alpha$.
Note that for $\alpha\sim 1$ the critical temperature reaches
values $\sim c_s q_F$.
\begin{figure}[t]
\includegraphics[scale=0.23]{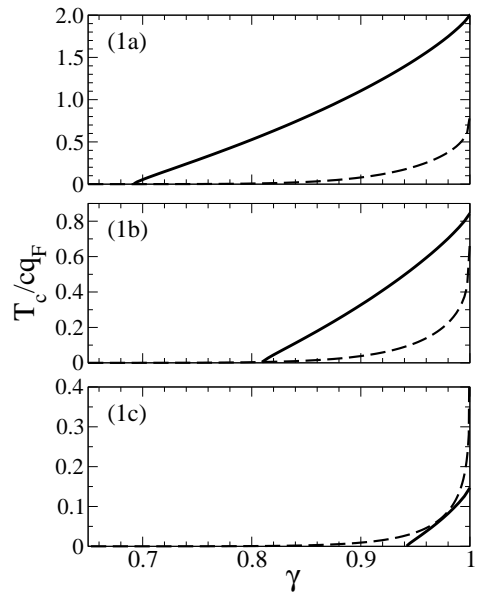}
\caption{ \label{cap1} (a) thru (c): Critical temperatures vs. fermion-phonon
coupling $\gamma$ for $\alpha =2 q_F\xi=1.0$, $1.4$, and $2.0$.  The solid line
is the critical temperature for $s$-wave odd-frequency pairing and the dashed
line for $p$-wave pairing. In all cases we have set $Z=1$. The $x$-axis is the same for all three
panels, but note the different $y$-axis scales.}
\end{figure}

For the even frequency p-wave case, we consider the effective potential
in the $l=1$ channel, using ${\cal D}_{1}$.  We note that this
is the same as the p-wave case studied in Refs.~\cite{bulgac}
except that we use the renormalized ${\cal D}_1$ following
from Eq.~(\ref{phonon})---this places the p-wave on equal footing for
comparison to the the s-wave odd frequency phase.
The p-wave phase exists as the coupling $\gamma$ goes to zero,
although the critical temperature is exponentially suppressed.
The p-wave critical temperature is plotted in Fig.~1(a-c)
for different values of the parameter $\alpha$ as a dashed line. Note
that due to the relative smallness of the effective coupling
strength ${\cal D}_1(0)$ the critical temperature for the p-wave
pairing is much lower than that for the odd frequency s-wave
(except for Fig.~1(c), where the odd-frequency region is very
small, i.e., for large $\alpha$).
\section{Possible Experimental Realization and Discussion}
Since most of the standard (in solid state) thermodynamic and transport
measurements are presently unavailable in cold atoms,
detecting odd-frequency pairing in a cold atom fermion-boson mixture
may be a non-trivial problem.
One intriguing possibility would be to take advantage of the
cold atom time-of-flight type experiments to study the unique correlations
of the odd-frequency phase.  For example, it has
been demonstrated in Ref.~\cite{greiner} that momentum correlations of
atomic fermions can be observed by the photodissociation of molecules;
upon release from a trap, the atoms exhibit density correlations between points ${\bf r}$
and $-{\bf r}$ (relative to the center of the trap) as a consequence of
the initial molecular state.  Similar measurements have been proposed
to detect other types of many-body correlations \cite{altman};
for example, a fermionic gas in the BCS regime has a
density-density correlation function $\langle n_F({\bf r}_1,t)n_F({\bf r}_2,t)\rangle$
which at sufficiently large times of flight $t$ is also peaked at
${\bf r}_1=-{\bf r}_2$ as a result of the pairing.

The above argument should, however, be modified when applied to
the odd-frequency type of pairing. Indeed, since the equal time
anomalous correlator is identically $0$ for the odd-frequency
pairing, there is no ${\bf r}_1=-{\bf r}_2$ correlation in
the two-point fermion correlation function. However, the
three-point correlation function $\langle n_F({\bf r}_1)n_F({\bf
r}_2)n_B({\bf r}_3)\rangle$, where $n_B$ is boson density, {\it
does} contain the signature of the odd-frequency pairing. To see
this one should notice that while $\langle \psi_{F{\bf
q}}(0)\psi_{F-{\bf q}}(0)\rangle =0$ for such pairing, $\langle
\psi_{F{\bf q}}(0){\dot \psi_{F-{\bf q}}(0)}\rangle\neq 0$.
Then, using $i{\dot \psi_F} = [H,\psi_F]$, where $H$ is given by
Eq.~(\ref{ham}), it is easy to show that $\langle\psi_{F{\bf
q}_1}(0)\psi_{F{\bf q}_2}(0)n_{B{\bf
q}_3}(0)\rangle\sim\delta ({\bf q}_1+{\bf q}_2+{\bf q}_3)$,
where $n_{B{\bf q}}$ is Fourier component of boson density and the width
of $\delta$ is primarily controlled by the size of the trap.
The boson density can be expressed as $n_{B{\bf q}}
\approx\sqrt{\rho_0}(\psi_{B{\bf q}}+\psi_{B-{\bf q}}^\dag)$, and
therefore the three point correlation function contains an
irreducible contribution peaked at ${\bf q}_1+{\bf q}_2=-{\bf
q}_3$ (in Fourier space).  This three-point correlation
can be interpreted as an order parameter first proposed in the context of
odd-frequency superconductivity in a t-J model \cite{Sasha}.
As a result the real space equal-time
correlation function $\langle n_F({\bf r}_1)n_F({\bf r}_2)n_B({\bf
r}_3)\rangle$ has a correlation peak at ${\bf r}_1+{\bf r}_2=-(m_B/m_F){\bf r}_3$,
where we have used the relationship between the wavevectors
of the particles in the initial state with their coordinates in the time-of-flight
image, ${\bf q}_{F,B}=m_{F,B}{\bf r}/t$; see Fig. 2. Therefore particle density cross-correlations which can, in principle, be deduced from instantaneous fermion and boson atom absorption images would provide a direct test for odd-frequency pairing in cold-atom mixtures.
\begin{figure}[t]
\includegraphics[scale=0.28]{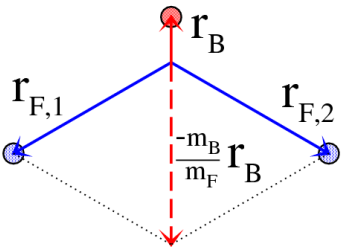}
\caption{ \label{cap2}  Schematic picture of three-particle correlations
under time-of-flight expansion: two fermions, $r_{{\rm F},1}$ and $r_{{\rm F},2}$,
and one boson, $r_{{\rm B}}$. }
\end{figure}

Finally we point out that the validity of Eq.~(\ref{gapeqn}) is controlled by the
Migdal criterion, i.e., the smallness of the vertex correction part.
The latter can be estimated as the three-legged diagram \cite{agd},
with two fermion and one phonon lines. The standard
order of magnitude estimate for the ratio between the bare and
the renormalized interaction vertices gives
$(\gamma/\sqrt{1-\gamma})\times(c_s/v_F)\times (q_F\xi)^{-1}$,
where we have used the phonon Green's function given by Eq.~(\ref{phonon}).
Thus, Eq.~(\ref{gapeqn}) is valid for finite $\gamma$, provided that
$\gamma$ is not too close to $1$, i.e., the point of the phase separation,
and $(c_s/v_F)\times (q_F\xi)^{-1}\ll 1$. Moreover, since
we are interested in the regime $q_F\xi\sim 1$, the vertex correction is small as long as $c_s/v_F\ll 1$.
Therefore the above results are quantitatively valid when $\xi_0 c_s m_F\ll 1\ (\hbar)$,
that is $m_F/m_B\ll 1$. While this condition is naturally
fulfilled in solid-state systems, with $m_F$ and $m_B$ being electron and ion masses,
this is not necessarily the case in cold atom systems. A reasonable choice
for testing our theory in trapped cold atom systems is a $^6$Li-$^{87}$Rb
binary mixture so that the mass ratio condition is satisfied.
Using the $^{87}$Rb background scattering length of $5.32$nm and a $^6$Li density of $10^{12}$cm$^{-3}$,
the coupling $\gamma=1$ corresponds to a $^{87}$Rb-$^{6}$Li scattering length of $22.7$nm, an order
of magnitude accessible via an interspecies Feshbach resonance.  To optimize the odd-frequency
phase, one would want $\alpha = 2 q_F \xi \sim 1$ (see Fig.~1).  With the same parameters as above,
$2 q_F \xi_0 =1$ corresponds to a $^{87}$Rb boson density of $2.3\times10^{14}$cm$^{-3}$ and the temperature
scale $T_c\sim c_s q_F \simeq 85 {\rm nK}$, readily accessible in cold atomic gases.

In summary, we have studied a mixture of bosons and single-species-fermions, showing that fermionic
superfluidity of the Berezinskii odd-frequency type is likely to exist under appropriate conditions,
i.e., this pairing occurs with a finite critical strength of the boson-fermion coupling.
We have estimated the transition temperature and pointed out the unique boson-fermion cross-correlations
which such a state exhibits.
\section{Acknowledgements}
We thank E. Abrahams, I. Kolokolov, V. V. Lebedev, I. Martin and E. Timmermans for valuable discussions.  The work is supported by the US DOE.
\section{Appendix}
To evaluate the threshold coupling and the critical
temperatures it is convenient to transform the ``Schrodinger''
equation ${\hat H}_l{\cal F}_l=\epsilon_{\infty,l} {\cal F}_l$ back to the
frequency representation, e.g., Eq.~(5), but with the
summation over the Matsubara frequencies replaced by integration
and with an additional $\epsilon_{\infty,l} {\cal F}_l$ term on the l.h.s.
Introducing $\Delta_l (\omega, \delta q) =
[\omega_n^2+{v_F^\ast}^2\delta q^2-\epsilon_{\infty,l}]
{\cal F}_l(\omega_n, \delta q)$, integrating out the momentum $\delta q'$ on the
r.h.s., and noticing that the solution $\Delta_l$
is independent of $\delta q$ (a result of evaluating
the phonon propagator on the Fermi surface), we obtain
\begin{equation}
\tag{A1}
\label{gap2}
\Delta_l(\omega)={(\lambda^\prime Z q_F)^2\over 4\pi v_F^\ast} \int\frac{d\omega^\prime}{2\pi}
{\Delta_l(\omega^\prime){\cal D}_l(\omega^\prime-\omega)\over \sqrt{{\omega^\prime}^2-\epsilon_{\infty,l}}}.
\end{equation}

We consider the $l=0$ and $l=1$ solutions to
Eqn.~(\ref{gap2}).  Since the effective coupling strength
${\cal D}_l(0)$ rapidly decreases for greater $l$, phases with higher
orbital momentum of the order parameter have much lower critical
temperatures and thus are never realized.

To estimate the critical coupling strength for the $l=0$ odd-frequency
solution let us first consider the zero temperature case ($\epsilon_{\infty,0}=0$ in
Eq.~(\ref{gap2})). Since $\Delta_0(\omega)$ must be odd, we use the ansatz
$\Delta_0(\omega)\propto\omega$ (with cut-off $\omega_c\sim c_s
q_F$); we have verified that this linear $\omega$-dependence as $\omega\rightarrow 0$ is
correct by means of explicit numerical solutions of the gap equation.
Expanding the r.h.s. of Eq.~(\ref{gap2}) linearly in $\omega$, canceling
$\omega$ and integrating the resulting r.h.s. of Eq.~(\ref{gap2}) by parts
we obtain the condition for the threshold (critical) coupling:
$4\pi^2 v_F^\ast \simeq (\lambda_c^\prime Z q_F)^2{\cal D}_0(0)$ (here we
have neglected the cut-off dependence assuming that
${\cal D}_0(\omega_c)\ll {\cal D}_0(0)$). It is instructive to evaluate the critical
coupling strength for $D(0) = D^0(0)$, that is, without accounting
for the phonon-mode softening. A straightforward calculation yields
$\gamma_c = {\lambda_c^\prime}^2 q_F^2/(2\pi^2\lambda v_F) =
\alpha_0^2/[Z^2\ln{(1+\alpha_0^2)}]$, where $\alpha_0 = 2q_F\xi_0$. Since
this $\gamma_c$ is always greater than $1$ (note that $Z\leq 1$),
one would conclude that the coupling needed for the formation of
the Berezinskii phase is stronger than that of the
phase separation ($\gamma=1$) and, therefore, that the
phase does not exist.  This conclusion, however, is erroneous because at
finite fermion-phonon coupling the renormalization of the phonon propagator, e.g.,
Eq.~(3), is crucial: as the coupling strength approaches
that of the phase separation, the effective interaction between
fermions increases due to phonon softening.  Thus, with $D(0)$ given by Eq.~(3)
we find that $\gamma_c$ satisfies the equation $\alpha^2/(Z^2\gamma_c)
= \ln{[1+\alpha^2/(1-\gamma_c)]}$ where $\alpha = 2q_F\xi$, which has solutions $\gamma_c\le 1$.
In particular, for $\alpha\gg1$ one finds
$\gamma_c\approx 1 - \alpha^2\exp(-\alpha^2/Z^2)$, and so the width
of the superfluid phase (in terms of the dimensionless coupling
$\gamma$) is exponentially small. For $\alpha\sim 1$,
$1-\gamma_c$ is {\it finite} and thus the superfluid phase exists only
when the effective boson coherence length $\xi$ is comparable with the
interfermion distance.
The transition temperature can be estimated
within the same linear ansatz for
$\Delta_0(\omega)$ by retaining the $-\epsilon_{\infty,0}$ in the
denominator on the r.h.s.~and solving the resulting equation
numerically for $\sqrt{-\epsilon_{\infty,0}}\sim T_{c,0}$, for
which the results are presented in Fig.~1(a-c) for different values of $\alpha$.

For the $l=1$ even-frequency p-wave phase, to estimate the
critical temperature $T_{c,1}$ or $\epsilon_{\infty,1}$ from
Eq.~(\ref{gap2}) it is sufficient to set $\Delta_1(\omega)={\rm const}$
(again, with cut-off $\sim c_s q_F$) and replace ${\cal D}_1(\omega-\omega^\prime)$ in
the r.h.s.~of Eq.~(\ref{gap2}) by ${\cal D}_1(0)$. After
a straightforward calculation using the phonon propagator of Eq.~(3) we obtain
that $T_{c,1}\sim c_s q_F \exp{(-1/g_1)},$
\begin{equation}
\tag{A2}
\label{pwave} g_1=
\frac{\gamma}{4\alpha^2}\Big{[}\Big{(}\frac{1-\gamma}{4\alpha^2}
 +\frac{1}{2}\Big{)}\ln(1+\frac{4\alpha^2}{1-\gamma})-1\Big{]},
\end{equation}
where, for simplicity, we have set $Z=1$. We note that the
p-wave phase has been considered in Refs.~[14] for fermion-boson
mixtures, and that our calculation would have been the same as those
if we had not used the renormalized phonon propagator.
The p-wave transition temperature given by Eqn.~(\ref{pwave})
is plotted in Fig.~1(a-c)
for different values of the parameter $\alpha$ as a dashed line.


\begin{thebibliography}{10}
\bibitem{comment3}
In Eq.~(\ref{phonon}) we have neglected the Landau damping term, which
is proportional to the ratio $c_s/v_F$, assumed to be small; see below.
\end{thebibliography}
\end{document}